\font\fthreei=cmti10 scaled\magstep1

\font\fthreeb=cmbx10 scaled\magstep1

\def\ii{\'\i}
\def\|{\'\i}

\def\lsim{\mathrel{\rlap{\lower4pt\hbox{\hskip1pt$\sim$}}
     \raise1pt\hbox{$<$}}}         %less than or approx. symbol
 \def\gsim{\mathrel{\rlap{\lower4pt\hbox{\hskip1pt$\sim$}}
     \raise1pt\hbox{$>$}}}         %greater than or approx. symbol
\def\ii{\' \i }

\def\na{-\kern-.4em\raise.8ex\hbox{{\tt \scriptsize a}}\ }
\documentclass[11pt]{article}

\def\today{\number\day \space de \ifcase\month\or
  janeiro\or fevereiro\or mar\c co\or abril\or maio\or junho\or
  julho\or agosto\or setembro\or outubro\or novembro\or
  dezembro\fi \space de \space
  \number\year}
\begin{document}

\centerline {\fthreeb A note on space dimensionality constraints relied on}
\vskip 10 pt
\centerline {\fthreeb  Anthropic arguments: methane structure and the origin of life}
\vskip 1.0 cm
\centerline {\fthreei F. Caruso $^{1,2}$}
\bigskip
\centerline {$^1$ {Centro Brasileiro de Pesquisas F\|sicas}}
\centerline {Rua Dr. Xavier Sigaud 150, 22290-180, Rio de Janeiro, RJ, Brazil}
\smallskip
\centerline {$^2$ {Instituto de F\|sica Armando Dias Tavares da Universidade do Estado do
Rio de Janeiro}}
\centerline {Rua S\~ao Francisco Xavier 524, 20550-013, Rio de
Janeiro, RJ, Brazil}
\medskip
\centerline{email: francisco.caruso@gmail.com}
\vskip 1.5 cm

\begin{flushright}
\begin{minipage}{7.2cm}
\baselineskip=8.pt
{\it Perhaps it is true that the hope that physical research can resolve the philosophical prob\-lems of space is just as vain as the hope that philosophical thought can resolve the physical problems of space.}
\vskip 0.5 cm
\hfill Max Jammer
%{\cite{Bachelard90}}
\end{minipage}
\end{flushright}

\vskip 1.0 cm

\baselineskip=12 pt

%\baselineskip=17 pt

%\noindent {\fthreeb 1.} {\fthreeb Introduction}
%\medskip

 Anthropic arguments have been proposed, independently, by philosophers and scientists to explain why we perceive a three dimensional Universe [1]. Some of them will be briefly reviewed in this note and a possible relationship between methane structure, the origin of life and space dimensionality will be pointed out.

\vspace{2mm}

Kant's conjecture~[2] that space three-dimensionality may, in some way, be related to Newton's inverse square law of Gravitation was the first step in this direction. Even though it has been shown [3] that Kant did not actually succeed in proving this conjecture -- indeed, he just concluded that there should be a relationship between this law and {\it extension} --, his contribution has the very merit of suggesting that the problem of dimensionality can also be treated in the framework of Physics and does not belong exclusively to the domain of Mathematics, neither to that of pure philosophical speculation. A deeper comprehension of Kant's conjecture had to wait the rise of field theory.

\vspace{2mm}

As a second step, one can quote the work of William Paley~[4], which can be considered the first attempt to shed light on the space dimensionality problem clearly from Anthropic arguments. In his work, Paley analyzes the consequences of changes in the form of Newton's gravitational law and of the stability of the solar system on human existence. Starting from a teleological thesis, his speculations take into account a number of mathematical arguments for an anthropocentric design of the World, which rest all upon the stability of the planetary orbits in our solar system and on a Newtonian mechanical {\it Weltanschauung}, as should be expected at that time.

\vspace{2mm}

In the twentieth century, the idea of how space dimensionality follows from the stability of planetary orbits in the solar system was revisited in Ehrenfest's seminal papers~[5-6]. Ehrenfest discusses several physical phenomena, where qualitative differences between three-dimensional $(\Re^3)$ and other $n$-dimensional $(\Re^n)$ spaces are found, such as the existence of stable planetary orbits and the stability of atoms and molecules. These aspects, which distinguish the $\Re^3$ Physics from the $\Re^n$ one, are called by him ``singular aspects" and his works were aimed at stressing them. A crucial assumption is built in the main ideas contained in~[5], namely that it is possible to make the formal extension $\Re^3 \rightarrow  \Re^n$ for a certain law of Physics and, then, find one or more principles that, in conjunction with this law, can be used to single out the proper dimensionality of space. For this approach to be carried out, in general, one has to decide how such formal extension will be done. Frequently, the form of a differential equation -- which usually describes a particular physical phenomenon in a three-dimensional space -- is preserved and its validity for an arbitrary number of dimensions is {\it postulated}. Then, ``singular aspects"\, may be brought into evidence by analyzing the higher dimensional mathematical solutions. For example, the Newtonian gravitational potential for a $\Re^n$-space, $V(r) \propto r^{2-n}$, is the solution of the generalized Laplace-Poisson equation,

$$ \sum\limits^{n}_{i=1} {{\partial^2 V}\over {\partial x_i^2} }= k \rho,$$

\vspace{2mm}

\noindent in an $n$-dimensional space. Based on the general solution of the above equation, assumed to correctly describe planetary motion in a space with $n$ dimensions, Ehrenfest has postulated the stability of orbital motion under central forces in order to constraint the number of dimensions. This general  procedure was also followed by Whitrow~[7]. Tangherlini~[8] noted that this approach could be broaden by proposing that for the Newton--Kepler problem, generalized to $\Re^n$ space, the principle to determine the spatial dimensionality could be summarized in the postulate that {\it there should be stable bound states orbits} -- or ``states" -- for the equation of motion governing the interaction of bodies, treated as material points. This will be generically called, from now on, the {\it stability postulate}. In his first paper~[8.a], Tangherlini showed that the essential results of the Ehrenfest--Whitrow investigation are unchanged when Newton's gravitational theory is replaced by General Relativity. Application of this same idea to the stability of hydrogen atom, described by a generalized Schr\"odinger equation, leads to the same kind of constraint in a very huge and different spatial scale.

\vspace{2mm}

In its essence, Ehrenfest's approach for planetary motion relies on two postulates: a) Poisson equation for any space dimensionality correctly explains the same phenomenon it describes in three dimensions; and b) the stability of the mechanical orbits should hold in the higher dimensional space. For him the former is the {\it causa formalis} and the later, the {\it causa efficiens} of space dimensionality. Actually, both are typical ingredients of any Anthropic constraint imposed on dimensionality. In spite of the fact that this kind of approach strongly reflects the recognition
of our ignorance being complete and assumes a `Principle of Similarity' -- using the expression adopted in [1], namely that alternative physical laws should mirror their actual form in three dimensions as closely as possible --  it seems a very hard task to avoid it as long as dimensionality is to be understood in the realm of Physics.

\vspace{2mm}

In any case, the previous results can be summarized by saying that only in universes in which gravity abides by an inverse square law could the solar system remain in a stable state over long time-scales. We will turn back to this point but, at this stage, it is important to stress that some epistemological and methodological aspects of this general approach based on the {\it stability postulates} were criticized in~[9].

\vspace{2mm}

This briefly reviews how the {\it stability postulate} is used to cast some light on
the problem of spatial dimensions.

\vspace{2mm}

It is important to stress that there is a third and decisive ingredient explicitly required or implicitly assumed every time a method which effectively connects the number of dimensions to some physical property is suggested. This is actually the most delicate part of any method one can propose for discussing the problem of spatial dimensions, and it will be shown that it is invariably connected to some version of Anthropic principle.

\vspace{2mm}

From the beginning, we would like to say that we are convinced that it is impossible to disentangle questions concerning this subject from some (any) kind of formalism representing a physical law, just because, as Jammer put it clearly [4],
``{\it ... it is clear that the structure of the space of Physics is not, (...), anything given in nature or independent of human thought. It is a function of our conceptual scheme.}"

\vspace{2mm}

This means that we should accept that the physical concepts and the concept of reality itself
acquire sense only within a theoretical construction where they can be discussed and realized. So far as the problem of space dimensions is considered, we must carefully examine the consequences of this fundamental point for the obvious fact that looking back to the History of Physics we soon realize that all theories and systems were built up, as Bertrand Russell said, assuming that ``{\it the limitation of the dimensions to three is {\rm (...)} empirical.}" [10]. Although this point has, in fact, motivated several works on the problem of spatial dimensions, it constitutes itself, at the same time, one of the main difficulties for discussing it, because the three-dimensionality of space is never questioned {\it a priori} when the physical law which is considered as the starting point for the search of any ``singular aspect" is established.
%In a certain sense, this essential difficulty could be bypassed if we are able to prove the validity of some other specific physical law whatever the number of spatial dimensions is, rather than simply postulating its general validity [9].

\vspace{2mm}

Even taking this criticism into account, a review of the literature on this subject leads us to regard to the Anthropic Principle as {\it an almost unavoidable} approach to the problem of space dimensionality when we want to explain why dimensionality is three and not another number. In any case, this is essentially related to Jammer's idea just recalled above. Thus, to the best of our knowledge, this epistemological limitation seems to be inherent to this problem (so far as we understand it) and, in a certain sense, is well illustrated and justified by the following Grassmann's words:

\begin{quotation}
{\noindent \it ``The concept of space can in no way be produced by thought,
but always stands over against it as a given thing. He who tries to
maintain the opposite must undertake the task of deducing the necessity of
the three dimensions of space from the pure laws of thought, a task whose
solution presents itself as impossible."} [11].
\end{quotation}

 These words just reinforce our conviction that the structure of physical space -- in particular its dimensionality -- is a function of our conceptual scheme and that it does not seem possible to formally deduce space dimensionality from it. In the last analysis, therefore, one should resort to phenomenology to determine it, which, at the end, actually means to accept some kind of Anthropic argument.

\vspace{2mm}

Let us turn back to Whitrow's argument. In his important 1955 paper, he asseverates that for trying ``{\it to isolate three-dimensional space as the only possibility for the world in which we find ourselves, we must now invoke some argument for showing why the number of dimensions cannot be {\rm less} than three}". To do this, he adapted the well known topological result from knot theory, that we cannot make a knot in even-dimensional space, to the necessity of higher forms of animal life to have brains in which electrical pulse informations carried on by nerves could not interfere destructively, which excludes a twofold and other even-fold spaces. This argument automatically constrains space to have an odd dimensionality $\geq 3$. Then, in the conclusion of this paper one can read:

\begin{quotation}
{\noindent \it ``Despite various recent attempts to show that [space dimensionality] is either a necessary attribute of our conception of physical space or is partly conventional and partly contingent, the problem cannot be considered as finally solved. A new attempt to throw light on the question indicates that this fundamental topological property of the world may possibly be regarded as partly contingent and partly necessary, since it could be {\rm inferred} as the unique natural concomitant of certain other contingent characteristics associated with the evolution of the higher forms of terrestrial life, in particular of Man, {\rm the formulator of the problem}."} [7.b].
\end{quotation}

Following a different approach, based on the {\it stability problem} and the Uncertainty Principle, Barrow \& Tipler [1] stressed that

\begin{quotation}
\noindent ``{\it (...) it has been claimed that if we assume the structure of the laws of Physics to be independent of the dimension, stable atoms, chemistry and life can only exist in $N<4$ dimensions.}".
\end{quotation}

And therefore they conclude, perhaps inspired on the aforementioned Whitrow's words, that ``{\it the dimensionality of the Universe is a reason for the existence of chemistry and therefore, most probably, for chemists also."}.

\vspace{2mm}

There is no doubt that both conclusions above have an Achilles' heel: there is indeed no support to the hypothesis that the laws of Physics are in principle independent of the dimensions, except simplicity; an example of application of Ockham's maxima {\it Entia non sunt multiplicanda praeter necessitatem}. Probably, in general, they are not. Let us remember, for example, that the group structure of the Euclidean Group of rotations is different for various numbers of dimensions. This fact has led Hermann Weyl to consider, in 1949, that ``{\it mathematical and physical laws may cease to be indifferent to the number of dimensions on some deeper level than has been touched by physics}" [12]. The result published br R.~Mirman, in 1984, that standard assumptions about the basic principles of Quantum Mechanics are not compatible with space-times with dimensions different from $(3+1)$, is also to be recalled [13]. In any case, it still remains the possibility that a particular physical system or dynamical process could have place in other dimensions but being described by a new mathematical law, in such a way that their main features and properties are maintained. The third objection could rise in the light of a 1999 paper for it contradicts all previous results based on the stability of atomic orbits, since the authors have claimed that there could have be a stable hydrogen atom in higher dimensions [14]. In addition they sustain that some spectroscopic experiment can be used to explain that our space is three-dimensional. However, such a result is based on very strong assumptions; for example, that ``{\it the specific expression for the force between charged particles and the stability of atoms are of more basic physical importance than the validity of Gauss' law"} and as a consequence Maxwell's equations should be modified. Criticisms of those ideas will appear elsewhere. In any case, despite of them, the intrinsic limitations of being still an approach which depends on Anthropic assumption could not yet be avoided. To the best of our knowledge, this is always the case.
Therefore, let us try to push it on by presenting now some remarks about the time (and space) ``scale" of the arguments previously discussed.

\vspace{2mm}

The first is related to Ehrenfest's stability argument which is typically valid for distances of the order of the solar system and in a time scale large enough to make the evolution of life possible on Earth, as mentioned by Whitrow [7].  However,  his argument about this subject [7.b] could be improved by stressing that it is not sufficient that the intensity of solar radiation on Earth's surface should not have fluctuated greatly for life still exist on
Earth; actually, the fact that Sun's spectra of radiation did not fluctuate very much should also be required [9].
By other side, Tangherlini's work about the stability of {\tt H} atoms is often invoked to suggest the validity of Chemistry in the same time scale as a necessary, although not sufficient, condition -- at least Chemical Thermodynamics of irreversible process should be also valid. Thus, as pointed out in [9],
``{\it the presence of atomic spectra in remote stars may also indicate[s] that space has had the same dimensionality at cosmic scale.}"
The existence of such a cosmic constraint on space dimensionality is a very interesting consideration and this subject was treated in [15].

\vspace{2mm}

The second one is also related to the general idea that among a large number of possible universes, the actual Universe is the one that which contains intelligent life, or at least had some form of life in a very long time scale. We have quoted above what Withrow, Barrow and Tipler said about human life and how it imposes some constraints on the number of dimensions. Infallibly this query addresses us to Biochemistry. There is a nice chapter on this subject on Barrow and Tipler's book [1], where several relevant topics are discussed in details,  and so will not be treated here. Among them we can quote the unique properties of carbon, hydrogen, oxygen and nitrogen, or whether or not it is possible to base life on elements other than these ones, and finally that those unique properties are probably necessary to guarantee the ecological stability required by highly-evolved life, although not sufficient. Our aim here is to introduce a new argument in favor of a stable scenario for space dimensionality for a time scale longer than that required for the existence of human or another kind of highly-evolved life on Earth, remembering that the usually accepted scales are: 2 Millions years ago the {\it homo erectus} have appeared, while the first {\it skeletons and easily recognizable fossils} range are of 600 Millions of years ago. This new argument is related to the methane structure as will be shown now.

\vspace{2mm}

Let us consider the famous experimental result published in 1959 by Harald C. Urey and Stanley Miller [16]. They showed to be possible, by means of an electrical discharge, to transform an admixture of gases consisting of methane, water, ammonia and hydrogen into a great number of organic compounds, among them some amino acids essential to life. Although it is not a {\it proof}, this result is widely considered as a strong evidence for the creation of life in a kind of primitive Earth atmosphere, quite different from that of the present days, composed of the four elements just mentioned. Accepting this means to accept that in a certain sense methane, which has the most simple formula among the organic compound ({\tt CH}$_4$), is somehow related to the origin of amino acids that could build up primitive life. In addition, it is implicit in this reasoning that the atomic structure and chemical properties of the elements have not changed.

\vspace{2mm}

Based on X-ray spectroscopy and on the empirical fact that an isomer of methane was never found, the tetrahedral structure of carbon was established [17]. In other words, Nature seems to have chosen just one spatial disposal for methane atoms and also for all compounds of the type $\mbox{\tt CH}_3\mbox{\tt Y}$ e $\mbox{\tt CH}_2\mbox{\tt YZ}$, with {\tt Y} and {\tt Z} being any group of atoms. This rules out any flat configuration for the simplest organic compound and requires, obviously, that the space in which it exists should be at least three dimensional.

\vspace{2mm}

So, to believe on Urey-Miller's experiment as a clue for the origin of amino acids essential to life, associated to an atmosphere possibly rich on methane, implicitly assumes that three is the minimum space dimensionality required by methane structure and for life to be developed this way. Putting this together with what was said above about the spectra of remote stars, a scenario where space dimensionality should be at least three for very large spatial and temporal scales seems plausible; much greater than that required by human life on Earth. Remember that some authors believe the origin of life -- probably thermophiles -- occurs $3,\!500$~Millions of years ago. Despite its speculative nature, this is a new constraint imposed not only on the number of dimensions but also on its stability throughout a very large space and time scale, obtained from a sort of modified strong Anthropic principle, namely, from the assumption that {\it the early Universe should necessarily contain amino acids}. A recent analysis [18] of the cosmic background radiation spectrum measured by COBE collaboration suggests that space dimensionality did not
vary significantly in a huge temporal scale, once this background radiation is expected to
be related to the Big Bang. This time scale can be safely put on the later epoch where the
universe was about $3\times 10^5$~yr old (red shift $z \simeq 10^3$). 

\vspace{2mm}

In conclusion, we would like to say that physicists and philosophers should still pay attention to many epistemological difficulties concerning the problem of space dimensionality, among which we could emphasize a certain incompleteness in the majority of approaches to this problem so far as they consider physical events taking place only in space, not in {\it space--time}. Thus, the problem of the number of space dimensions and that of time dimensions are probably not independent. Finally, whether or not a deeper comprehension on the problem of space dimensionality is to be reached and, in particular, if it could be possible to go on discussing this problem without taking into account any kind of Anthropic argument as some stage of a particular reasoning are still good questions without good answers.

\vskip 1.0 true cm

\noindent {\bf Acknowledgments}
\vskip 0.5 true cm
  We would like to thank the editor, L.C.B.~Crispino, for the kind invitation for contributing to J.M.F.~Bassalo's {\it Festschrift}, which we promptly accepted with great honor. We feel mandatory in this occasion to publicly acknowledge that our good friend Bassalo had a very huge influence on our earliest interest on History and Philosophy of Science, by reading many of his papers and essays. It is a pleasure to thank our friends Gilvan Alves, H\'elio da Motta and Mauro Velho de Castro Faria for reading the manuscript and for useful suggestions.

\vskip 1.0 true cm
%\newpage
\noindent {\bf References}
\medskip
\begin{description}
\item{[~1]} For a general introductory review see Barrow, J.D. \& Tipler, F., {\it The Anthropic Cosmological Principle}.
Oxford: Claredon Press, 1986.

\item{[~2]} Kant, I., {\it Gedanken von der wahren Sch\"atzung der
lebendigen Kr\"afte und Beurtheilung der Beweise, deren sich Herr von
Leibniz und andere Mechaniker in dieser Streitsache bedient haben, nebst
einigen vorhergehenden Betrachtungen, welche die Kraft der K\"orper
\"uberhaupt betreffen}, K\"onigs\-berg, 1747; reprinted in: Kant {\it
Werke}, Band 1, {\it Vorkritische Schriften}, Wissenschaftliche
Buchgesellschaft Darmstadt, 1983. English translation of part of this work
was done by J. Handyside and published as part of the book {\it Kant's
inaugural dissertation and the early writings on space}. Chicago: Open
Court, 1929, reprinted by Hyperion Press, 1979.

\item{[~3]} Caruso, F. \& Moreira Xavier de Ara\'ujo, R., ``Sull'Influenza di Cartesio, Leibniz e Newton nel primo Approccio di Kant al Problema dello Spazio e della sua Dimensionalit\`a", {\it Epistemologia} (Genova, Italia) v.~XXI, pp.~211-224 (1998). Se also from the same authors ``On Kant's first insight into the problem of dimensionality and its physical foundations", {\it Notas de F\ii sica CBPF}, NF-079/96 (1996). A slightly different version of the last paper was published in Portuguese as
    ``Notas sobre o problema da dimensionalidade do espa\c co e da extens\~ao no primeiro texto do jovem Kant",
{\it Scientia} (S\~ao Leopoldo) {\bf 7} (2), pp. 13-22 (1996).

\item{[~4]} Paley, W., {\it Natural Theology}, 1802, reprinted in {\it The Works
of William Paley}, 7 volumes, edited by R. Lynam, London, 1825. A recent book's reprint is available from Oxford University Press, 2006.

\item{[~5]} Ehrenfest, P., ``In what way does it become manifest in the
fundamental laws of physics that space has three dimensions?", {\it Proc.
Amsterdam Acad.} {\bf 20} (1917) 200; reprinted in  Klein M.J. (ed.),
{\it Paul Ehrenfest -- Collected Scientific Papers}. Amsterdam: North
Holland Publ. Co., pp.~400-409 (1959).

\item{[~6]} Ehrenfest, P., ``Welche Rolle spielt die
Dreidimensionalit\"at des Raumes in den Grund\-gesetzen der Physik?", {\it
Annalen der Physik} {\bf 61}, 440 (1920).

\item{[~7]}  Whitrow, G.J., {\it The Structure and Evolution of the
Universe}. New York: Harper and Row, 1959; and b) ``Why Physical Space has
Three Dimensions?", {\it Brit. J. Phil. Sci.} {\bf 6}, pp.~13-31 (1955).

\item{[~8]} Tangherlini, F.R., a) ``Schwarzschild Field in $n$ Dimensions
and the Dimensionality of Space Problem", {\it Nuovo Cimento}
{\bf 27}, 636 (1963). b) ``Dimension\-ality of
Space and the Pulsating Universe", {\it ibid} {\bf 91B}, 209 (1986).
%\item {[ 9]} Barrow J.D. \& Tipler F., {\it op. cit.}, p. 261.

\item{[~9]} Caruso, F. \& Moreira Xavier, R., ``On the physical problem of spatial dimensions: an alternative procedure to stability arguments", {\it Fundamenta Scienti\ae}
{\bf 8}, (1), pp. 73-91 (1987).

\item{[10]}  Russell, Bertrand, {\it An Essay on the Foundations of Geometry}. Cambridge:
University Press (1897).

\item{[11]} Grassmann, {\it Die Ausdehnungslehre}, 2nd. ed., Leipzig 1878,
p.~XXIII, as quoted by Jammer M., {\it op. cit.}, p. 184.

\item{[12]} Wey, Hermann, {\it Philosophy of Mathematics and Natural Science}, Princeton: University Press, 1949, p.~136.

\item{[13]} Mirman, R., ``The Dimension of Space-Time", {\it Lettere al Nuovo Cimento} {\bf 39} (16), pp.~398-400 (1984).

\item{[14]} Burgbacher, F.; Lämmerzahl, C.; and Macias, A., ``Is There a Stable Hydrogen
Atom in Higher Dimensions?",  {\it Journal of Mathematical Physics} {\bf 40} (2), pp.~625-634 (1999).

\item{[15]} Caruso, F. \& Moreira Xavier, R., ``Space dimensionality: what can we learn from stellar spectra and from the M\"ossbauer effect", invited paper {\it in} R.B. Scorzelli, I. Souza Azevedo \& E. Baggio Saitovitch (Eds.), {\it Essays on Interdisciplinary Topics in Natural  Sciences Memorabilia: Jacques A. Danon}. Gif-sur-Yvette/Singapore: \'Editions Fronti\`eres, pp.~73-84, 1997.

\item{[16]} Urey, H.C. \&  Miller, S., ``Organic compound synthesis
on the primitive earth". {\it Science} {\bf 130}, p.~245-251; ``Origin of Life''
(reply to letter by S.W.~Fox). {\it Science} {\bf 130}, pp.~1622-1624 (1959).

\item{[17]} For a didactical presentation see Caruso, F. \& Oguri, V., {\it F\ii sica Moderna: Origens Cl\'assicas e Fundamentos Qu\^anticos}. Rio de Janeiro: Elsevier/Campus, 2006.
    
\item{[18]} Caruso, F. \& Oguri, V., ``The Cosmic Microwave Background Spectrum and
an Upper Limit for Fractal Space Dimensionality", ArXiv: 0806.2675 [astro-ph] 17 June.

\end{description}
\end{document}